\documentclass[twocolumn,secnumarabic,amssymb, nobibnotes, superscriptaddress,aps, prd]{revtex4}

\newcommand{\bea}{\begin{eqnarray}}
\newcommand{\ena}{\end{eqnarray}}
\newcommand{\be}{\begin{equation}}
\newcommand{\ee}{\end{equation}}
\newcommand{\beann}{\begin{eqnarray*}}
\newcommand{\enann}{\end{eqnarray*}}
\usepackage[utf8]{inputenc}
\usepackage{graphics,graphicx}
\usepackage{amsmath,amssymb,amsfonts,latexsym,cancel}
\usepackage{bm} 
\usepackage{psfrag}
\usepackage{color}
\usepackage{multirow}
\usepackage{float} 
\usepackage{verbatim}
\usepackage{caption}
\usepackage{subfigure}

\begin{document}

\title{Massive neutron stars with holographic multiquark cores}

\author{Sitthichai Pinkanjanarod}
\email{quazact@gmail.com, Sitthichai.P@student.chula.ac.th}
\affiliation{High Energy Physics Theory Group, Department of Physics,Faculty of Science, Chulalongkorn University, Bangkok 10330, Thailand}
\affiliation{  Department of Physics, Faculty of Science, Kasetsart University, Bangkok 10900, Thailand}

\author{Piyabut Burikham}
\email{piyabut@gmail.com, piyabut.b@chula.ac.th}
\affiliation{High Energy Physics Theory Group, Department of Physics,Faculty of Science, Chulalongkorn University, Bangkok 10330, Thailand}

\date{\today}

\begin{abstract}
    Phases of nuclear matter are crucial in the determination of physical properties of neutron stars~(NS). In the core of NS, the density and pressure become so large that the nuclear matter possibly undergoes phase transition into a deconfined phase, consisting of quarks and gluons and their colour bound states. Even though the quark-gluon plasma has been observed in ultra-relativistic heavy-ion collisions\cite{Gyulassy, Andronic}, it is still unclear whether exotic quark matter exists inside neutron stars. Recent results from the combination of various perturbative theoretical calculations with astronomical observations\cite{Demorest, Antoniadis} shows that (exotic) quark matter could exist inside the cores of neutron stars above 2.0 solar masses ($M_{\odot}$)~\cite{Annala:2019puf}. We revisit the holographic model in Ref.~\cite{bch, bhp} and implement the equation of states~(EoS) of multiquark nuclear matter to interpolate the pQCD EoS in the high-density region with the nuclear EoS known at low densities. For sufficiently large energy density scale~($\epsilon_{s}$) of the model, it is found that multiquark phase is thermodynamically prefered than the stiff nuclear matter above the transition points. The NS with holographic multiquark core could have masses in the range $1.96-2.23~(1.64-2.10) M_{\odot}$ and radii $14.3-11.8~(14.0-11.1)$ km for $\epsilon_{s}=26~(28)$ GeV/fm$^{3}$ respectively. Effects of proton-baryon fractions are studied for certain type of baryonic EoS; larger proton fractions could reduce radius of the NS with multiquark core by less than a kilometer.
	
\end{abstract}
\maketitle


\section{Introduction}\label{sec-introd}
In the final fate, a star collapses under its own gravity when the internal pressure from nuclear fuel is depleted. The quantum pressure of fermions kicks in to rescue.  If the mass of the star is below $0.7$ solar mass, the degeneracy pressure of neutrons alone would be able to stop the collapse~\cite{tov1,tov2}. Effects of repulsive nuclear force help support the neutron star up to higher masses $>1.4 M_{\odot}$. When a star is more massive than the upper mass limit of the neutron star, it would collapse into a black hole eventually. However, there is a possibility that under extreme pressure and density, the quarks within hadrons would become effectively deconfined from the localized hadrons but still be confined by gravity within the star. The deconfined phase of quarks could generate larger pressure to sustain even more massive neutron stars or even quark stars. 

Even in the deconfined phase, quarks can still form bound states via the remaining Coulomb-like strong interaction mediated by unconfined gluons, the multiquark states. Observations of multiquark candidates such as pentaquark and tetraquark have been accumulated for decades, see e.g. Ref.~\cite{Aaij:2020fnh} for the latest report. It is only natural to imagine an abundance of multiquarks in the core of dense stars where the deconfined quarks are extremely compressed tightly close together. Due to the nonperturbative nature of the strong interaction, the difficulty of lattice QCD approach when dealing with finite baryon density, and a reliability issue of MIT bag as a tool to study the behaviour of the deconfined quarks and gluons in the dense star, we use the equation of state of the deconfined nuclear matter from the holographic model as a complementary tool to investigate the properties of the dense star. There are some studies on the holographic model of deconfined quark matter in the dense star e.g. in D3/D7 system~\cite{Ecker:2019xrw} and in D4/D8/$\overline{\text{D8}}$ system~\cite{bch,bhp}.

Recent work~\cite{Annala:2019puf} reveals potentially two effective power-law equations of states~(EoS) interpolating between low and high density EoS calculated from the Chiral Effective Field Theory~(CET)~\cite{Tews:2012fj} and perturbative QCD~\cite{Andersen:2011sf,Mogliacci:2013mca}. The empirical EoS gives adiabatic index and sound speed characteristic of the quark matter phase, showing evidence of quark core within the NS. In this work, we revisit the holographic model investigated in Ref.~\cite{bhp} and match the EoS of multiquark nuclear matter with the low and high density EoS and demonstrate that it can interpolate well between the two regions. The masses of NS with multiquark core are consistent with current observations, allowing NS with $M\gtrsim 2 M_{\odot}$~\cite{Demorest,Antoniadis}. Depending on the colour states of multiquark, the mass could be as high as $2.2-2.3 M_{\odot}$, still too light to be a candidate for the object recently found by LIGO/Virgo~\cite{Abbott:2020khf} which requires mass around $2.50-2.67 M_{\odot}$. 

This work is organized as the following. Section~\ref{sec-holomq} reviews holographic model studied in Ref.~\cite{bhp} and presents the EoS of multiquark nuclear matter. Section~\ref{sec-eosns} summarizes the EoS from CET and piecewise polytrope used in the interpolation and EoS of the multiquark core in the high density region. Thermodynamic analysis of phase transition between the baryonic matter and multiquark phase is discussed in Section~\ref{sectPT}. Mass-radius diagram, mass-central density relation and thermodynamic properties of NS with multiquark core are explored in Section~\ref{sec-mr}. Section~\ref{sec-con} concludes our work.

\section{Holographic multiquark and the EoS}\label{sec-holomq}
Within the framework of gauge-gravity duality from superstring theories, bound states of quarks in the boundary gauge theory can be described holographically by strings and branes.  Mesons can be expressed as a string hanging in the bulk with both ends locating at the boundary of the AdS space\cite{maldacena2} while baryons can be represented by D$p$-brane wrapped on the $S^{p}$ with $N_c$ strings attached and extending to the boundary of the bulk space\cite{witb,gross&ooguri}.  The gauge theory from the original AdS/CFT duality is still different from the actual gauge theory described by QCD. The gauge theory from gravity dual that captures most features of QCD is the Sakai-Sugimoto (SS) model\cite{ss lowE, ss more}. In this model, hadrons naturally exist in the confined phase however, another kind of bound states of quarks can also occur in the deconfined phase at the intermediate temperatures above the deconfinement, the multiquark states~\cite{bch,bhp}. See e.g. Ref.~\cite{Burikham:2011zz} for a concise review of holographic multiquarks.

\subsection{Holographic multiquark configuration}
The configuration in the SS model consists of D4-brane background and D8/$\overline{\text{D8}}$ flavor branes. $N_c$ D4-branes provides 4D SU($N_c$) Yang-Mills gauge theory holographically. On the other hand, $N_f$ D8/ $N_f$  $\overline{\text{D8}}$ flavor branes provide a description for confinement/deconfinement phase transition depending on the configuration of the branes. In terms of symmetry, $N_f$ D8/ $N_f$ $\overline{\text{D8}}$ flavor branes poses the global symmetries U$(N_f)_L$ and U$(N_f)_R$ which can fully describe U$(N_f)_L$ $\times$ U$(N_f)_R$ chiral symmetry breaking when the D8 and $\overline{\text{D8}}$ are connected. At low energy, the classical solution of the field configuration on the gravity side suggests a cigar-like shape for the compactified spatial direction of a confined background. At high temperature, the cylindrically compactified background spacetime with flavor branes in parallel embedding is preferred, therefore the broken chiral symmetry is restored and the corresponding nuclear matter phase becomes deconfined~\cite{Aharony_chiral}. 

\begin{figure}
	\centering
	\includegraphics[width=0.45 \textwidth]{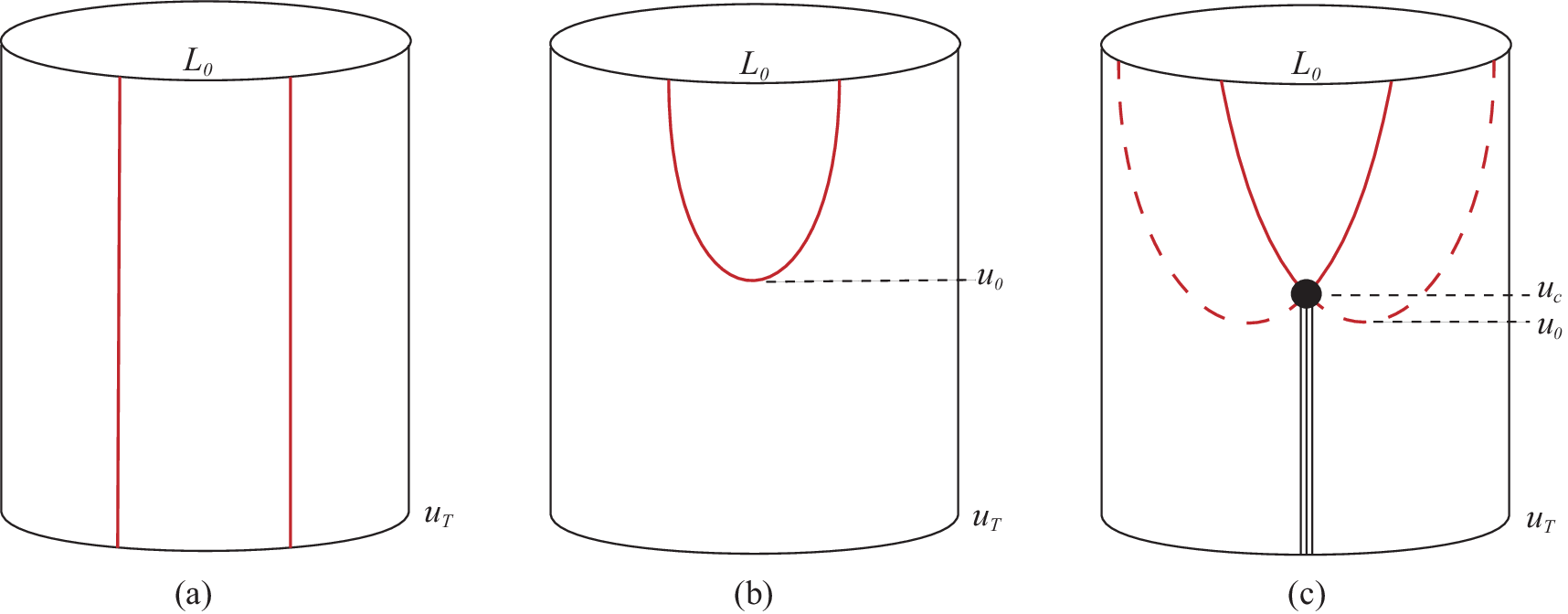} \caption{Different
		configurations of D8 and $\overline{\text{D8}}$-branes in the Sakai-Sugimoto model that are dual to the phases of
		(a) ${\chi}_S$-QGP, (b) vacuum and (c) multiquark phase.  The asymptotic separation of D8-branes is fixed to $L_0=1$.~\cite{bch} } \label{phase}
\end{figure}

In the deconfined phase~\cite{bch}, there are 3 possible configurations as shown in Fig.~\ref{phase}: (i)
the parallel configuration of both D8-branes and $\overline{\text{D8}}$ representing the
$\chi_S$-QGP~(chiral symmetric quark-gluon plasma) and (ii) connected D8-$\overline{\text{D8}}$ without sources in the bulk
representing the vacuum with broken chiral symmetry. Another stable configuration~(iii) is multiquark phase consisting of the connected D8-$\overline{\text{D8}}$ branes with the D4-brane as the baryon vertex submerged and localized in the middle of the D8 and $\overline{\text{D8}}$. The baryon vertex can be attached with radial hanging strings that represent colour charge of the multiquark configuration. To make connection with strong interaction we will set $N_{f}=3$, i.e., considering only light quark flavours, and focus on the common aspects of QCD and the holographic model in the large $N_{c}$ limit. 

\subsection{Equation of state}  \label{sec-eos}
Holographically, the grand canonical potential and the chemical potential of the multiquark matter are given by~\cite{bhp}

\begin{eqnarray}
\Omega &=& \int^{\infty}_{u_{c}}du {\displaystyle{\left[
		1-\frac{F^{2}}{f(u)(u^{8}+u^{3}n^{2})}\right]^{-\frac{1}{2}}}}
\frac{u^{5}}{\sqrt{u^{5}+n^{2}}},\; \label{eq:Grand}\\
\mu &=& \int^{\infty}_{u_{c}}du {\displaystyle{\left[
		1-\frac{F^{2}}{f(u)(u^{8}+u^{3}n^{2})}\right]^{-\frac{1}{2}}}}\frac{n}{\sqrt{u^{5}+n^{2}}}\nonumber\\
& &+ \frac{1}{3}u_{c}\sqrt{f(u_{c})}+n_{s}(u_{c}-u_{T}) \label{eq:mu}
\end{eqnarray}
\noindent respectively, where $u$ is a radial coordinate of the background metric of
the bulk spacetime in the SS model in a deconfined phase at finite temperature $T$, $f(u)\equiv 1-u_{T}^{3}/u^3,$ $ u_T=16{\pi}^2 R^3 T^2/9,$ $ R^3\equiv
\pi g_s N_c l_{s}^3,$ $l_s$ is the string length and $g_s$ is the string coupling. $u_{c}$ is the position of the baryon vertex source as shown in Fig.~\ref{phase}.  The asymptotic separation of D8-branes is fixed to $L_0=1$. $n_{s}$ is the number fractions of radial strings $k_{r}$ in the unit of $N_{c}$ that represents the colour charges of a multiquark configuration. $n(u)$ is the baryon number density which is a constant of the configuration given by

\begin{eqnarray}\label{d_const}
n(u)&=&\frac{u\hat{a}'_{0} }{\sqrt{f(u)(x'_4
		)^2+u^{-3}(1-(\hat{a}'_{0})^2)}}=\text{const.}
\end{eqnarray}
where $x_4$ is the compactified coordinate transverse to the probe D8/$\overline{\text{D8}}$ branes with arbitrary periodicity $2\pi R$. The $\hat{a} = 2
\pi \alpha^{\prime}\hat{A}/(R\sqrt{2 N_{f}})$ is a rescaled version of $\hat{A}$, the originally diagonal $U(1)$ gauge field, where $\alpha^{\prime}$ is a universal Regge slope. The position $u_{c}$ of the vertex is determined from the
equilibrium condition of the D8-D4-strings configuration~(see
Appendix A of Ref.~\cite{bch}). Another constant of the configuration is
\begin{eqnarray}\label{x4}
(x^{\prime}_{4})^{2}& = & \frac{1}{u^{3}f(u)}\Big[
\frac{f(u)(u^{8}+u^{3}n^{2})}{F^{2}}-1 \Big]^{-1}=\text{const.},
\label{eq:x4}
\end{eqnarray}
where $F$ is a function of $u_c$, $n$, $T$ and $n_s$, given by
\begin{equation}
F^{2} = u_{c}^{3} f(u_{c}) \left( u_{c}^{5}+n^2 -\frac{n^2
	\eta_{c}^{2}}{9 f(u_{c})}\right),
\end{equation}
where $\eta_{c}\equiv 1+\frac{1}{2}\left(\frac{u_T}{u_c}\right)^3
+3 n_s \sqrt{f(u_c)}$. 

Thermodynamic relations of multiquark states can be found in Ref.~\cite{bhp}. The grand potential $G_{\Omega}$ can be written as
\begin{equation}
dG_{\Omega} = -P dV -S dT-N d\mu
\end{equation}
where the state parameters $P$, $V$, $S$,
$T$, and $N$ are the pressure, volume, entropy, temperature, and the total
number of particles of the system respectively.  Since the change
of volume is not our main concern, we define the volume density of
$G_{\Omega}$, $S$ and $N$ to be $\Omega$, $s$ and $n$,
respectively. Therefore, we have, at a particular $T$ and $\mu$,
\begin{equation}
P=-G_{\Omega}/V \equiv -\Omega(T,\mu).
\end{equation}
Assuming that the multiquark states are spatially uniform, we
obtain
\begin{equation}
n=\frac{\partial P}{\partial \mu}(T,\mu).
\end{equation}
Using the chain rule,
\begin{equation}
\frac{\partial P}{\partial n}\Bigg\vert_{T}=\frac{\partial
	\mu}{\partial n}\Bigg\vert_{T}\; n,
\end{equation}
so that
\begin{equation}
P(n,T,n_s)=\mu(n,T,n_s)~n -\int_{0}^{n} \mu(n',T,n_s)
~\text{d}(n'),\label{pmud}
\end{equation}
where the regulated pressure is assumed to be zero when
there is no nuclear matter, i.e. $n=0$.

\subsubsection{Equation of state for multiquark}

In the limit of small $n$, the baryon chemical potential in Eqn.\eqref{eq:mu} can be
approximate as
\begin{eqnarray}
\mu & \simeq & \mu_{source} + \alpha_{0} n - \beta_{0}(n_s) n^3, \label{muofd}
\end{eqnarray}
where
\begin{eqnarray}
\mu_{source}& \equiv &\frac{1}{3}u_{c}\sqrt{f(u_{c})}+n_{s}(u_{c}-u_{T})\notag \\
\alpha_{0}& \equiv &\int_{u_0}^{\infty} du
\frac{u^{-5/2}}{1-\frac{f_{0}u_{0}^8}{fu^8}} ~, \notag \\
\beta_{0}(n_s)& \equiv &\int_{u_0}^{\infty} du
\frac{u^{-5/2}}{2\sqrt{1-\frac{f_{0}u_{0}^8}{f u^{8}}}}\nonumber\\
& &\times\left[\frac{f_0
	u_{0}^{3}}{fu^8-f_{0}u_{0}^{8}}
\left(1-\frac{\eta_{0}^{2}}{9f_0}-\frac{u_{0}^{5}}{u^{5}}\right)+\frac{1}{u^5}\right],\notag
\end{eqnarray}
and $u_{0}$ is the position when $x'_{4}(u_{0})=\infty$ as shown in Fig.~\ref{phase}.

By substituting Eqn.\eqref{muofd} into Eqn.\eqref{pmud}, the pressure in the limit of small $n$ can be expressed as
\begin{equation}
P\simeq \frac{\alpha_{0}}{2} n^2 -\frac{3 \beta_{0}(n_s)}{4} n^4.
\label{eq:Plow}
\end{equation}

In the limit of large $n$ and relatively small $T$,
\begin{eqnarray}
\mu & \approx & \mu_{source} + \frac{n^{2/5}}{5} \frac{\Gamma
	\left(\frac{1}{5}\right)
	\Gamma\left(\frac{3}{10}\right)}{\Gamma\left(\frac{1}{2}\right)}\nonumber\\
&&+\frac{u_{c}^{3} f_{c}}{10}
\left(1-\frac{\eta_{c}^{2}}{9f_c}\right)n^{-4/5} \frac{\Gamma
	\left(-\frac{2}{5}\right)\Gamma\left(\frac{19}{10}\right)}{\Gamma\left(\frac{3}{2}\right)}
\label{eq:muhd}
\end{eqnarray}
\noindent where the term from lower limit of integration in Eqn.\eqref{eq:mu}, $u_{c}^5/n^2$ approaches zero as $n$ becomes very large. Again by using Eqn.\eqref{pmud}, we obtain

\begin{equation}
P \simeq \frac{2}{35}\left( \frac{\Gamma\left(\frac{1}{5}\right)
	\Gamma\left(\frac{3}{10}\right)}{\Gamma\left(
	\frac{1}{2}\right)}\right) n^{7/5}.\label{eq:Phigh}
\end{equation}
Also the energy density can be found via the relation
$d\rho=\mu dn$ and the chemical potential is given by
\be
\mu = \int_{0}^{n}\frac{1}{\eta}\left( \frac{\partial P}{\partial \eta}\right)~d\eta +\mu_{0},
\ee
where $\mu_{0}\equiv \mu(n=0)$. The main results from Ref.~\cite{bhp} are summarized as
\bea
P &=& a n^{2}+b n^{4},  \notag \\
\rho &=& \mu_{0}n+a n^{2}+\frac{b}{3}n^{4},  \label{eosmq1}
\ena
for small $n$ and 
\bea
P &=& k n^{7/5},  \notag \\
\rho &=& \rho_{c}+\frac{5}{2}P+\mu_{c}\left[ \left( \frac{P}{k}\right)^{5/7}-n_{c} \right]\notag \\
&&+kn_{c}^{7/5}-\frac{7k}{2}n_{c}^{2/5}\left( \frac{P}{k}\right)^{5/7},  \label{eosmq2}
\ena
for large $n$ respectively. For $n_{s}=0$, it is numerically determined in Ref.~\cite{bhp} that $n_{c}=0.215443, \mu_{c}=0.564374$ for large $n$, and $a=1, b=0, \mu_{0}=0.17495$ for small $n$. For $n_{s}=0.3$ we have $n_{c}=0.086666, \mu_{c}=0.490069$ for large $n$, and $a=0.375, b=180.0, \mu_{0}=0.32767$ for small $n$. And $k=10^{-0.4}$ is valid for both cases reflecting universal behaviour at high density. $n_{c}, \mu_{c}$ are the number density and chemical potential where the EoS changes from large to small $n$. Notably, these coefficients and exponents of the power laws in the EoS are completely determined by the holographic SS model with only two free parameters, the number of hanging strings representing the colour charges $n_{s}$ and the energy density scale~\cite{bhp}. 

Notably as shown in Ref.~\cite{bhp}, the pressure~(as well as the chemical potential and density) of the multiquark matter is quite insensitive to the change of temperature in the range of values where the multiquark phase is thermodynamically prefered, $T<10^{12}$ K, the (chirally symmetric) QGP phase transition temperature. The position $u_{c}$ changes less than $0.4 \%$ in such temperature range~(shown in Fig.~2 of Ref.~\cite{bhp}). We can thus use the same multiquark EoS given above for all temperature less than $10^{12}$ K down to very low temperature.  We are only interested in modelling the deconfined phase via holography.  We will hence ignore the deconfinement-confinement phase transition in the Sakai-Sugimoto model and just extrapolate our results down to neutron star temperatures.  Since we are using the CET and other empirical EoS for the confined baryonic matter in this work, the phase transition analysis can be performed without considering the Hawking-Page transition between the confined and deconfined phase in the SS model.

\section{EoS of the NS} \label{sec-eosns}

The structure of neutron star can be investigated through observations and modeling of the strongly interacting hadronic matter for both the low-density crust and high-density core inside the star. However, with absence of the direct first-principle calculation at densities above the nuclear matter saturation (baryon number) density $n_0 \approx 0.16 ~{\rm fm}^{-3}$, an accurate determination of the state of matter inside the NS cores is still not possible. Recent observations start offering empirical constraints in both opposing low density and high density limits of the nuclear matter inside the NS. Therefore, not only the model-independent approach~\cite{Annala:2019puf} to the problem has become feasible but also it could provide a hint for the viable physical equation of states of nuclear matter inside the NS. 

\subsection{EoS of nuclear matter in low and intermediate density regime}
At low density, there is a limitation that comes from the well-studied NS crust region~\cite{Fortin} to the density $n_{\rm CET} \equiv 1.1n_0$, where matter occupies the hadronic-matter phase using chiral effective field theory (CET) which provides the EoS to good precision, currently better than $\pm 24\%$ \cite{Gandolfi,Tews:2012fj}. 

For very low density crust, EoS can be found from Table 7 of Ref.~\cite{Hebeler_2013}, it can be fit with the series of polytropes below,
\begin{eqnarray}
P(\rho) &=& \kappa_a \rho ^{\Gamma_a} + \alpha_a, \; \text{for} \; 0 \leq \rho \leq \rho_a,\notag \\
P(\rho) &=& \kappa_b \rho ^{\Gamma_b}, \; \text{for} \; \rho_a \leq \rho \leq \rho_b,\notag \\
P(\rho) &=& \kappa_c \rho ^{\Gamma_c}, \; \text{for} \; \rho_b \leq \rho \leq \rho_c,\notag \\
P(\rho) &=& \kappa_d \rho ^{\Gamma_d}, \; \text{for} \; \rho_c\leq \rho \leq \rho_d. \notag \\ \label{EoS7} 
\end{eqnarray}
where $(\kappa_a,\Gamma_a,\alpha_a)$ = (280.00, 2.0000, $-6.0000\times 10^{-21}$) and $(\kappa_b,\Gamma_b)$ =  ($2.15247\times 10^{-3}$, 1.22213), $(\kappa_c,\Gamma_c)$ =  ($9.08176\times 10^{-4}$, 0.622687), $(\kappa_d,\Gamma_d)$ = ($3.70286\times 10^{-4}$, 1.61786), while $(\rho_a,\rho_b,\rho_c,\rho_d)$ = ($2.66284\times 10^{-7}$, 0.237033, 2.46333, 75.1364) for the pressure and density expressed in the unit of $\text{MeV fm}^{-3}$ and $\text{MeV fm}^{-3}c^{-2}$ respectively.  

For slightly higher density in the range $75.1364~\text{MeV/fm}^{3}<\rho c^{2}<165.3~\text{MeV/fm}^{3}$ of Table 5 of Ref.~\cite{Hebeler_2013}, the energy density and pressure of the nuclear matter can be expressed as
\begin{eqnarray}
\rho(\bar{n})c^{2}/T_{0} &=& a_{0}\bar{n}^{5/3}+b_{0}\bar{n}^{2}+c_{0}\bar{n}^{\gamma +1},
\label{eq:E_Nucl}
\end{eqnarray}
where $\bar{n}=n/n_{0}$ and 
\begin{eqnarray}
P(\bar{n})/T_{0} &=& \frac{2}{3}n_{0}a_{1}\bar{n}^{5/3}+n_{0}b_{1}\bar{n}^{2}+\gamma n_{0}c_{1}\bar{n}^{\gamma +1},
\label{eq:P_Nucl}
\end{eqnarray}
for $T_{0}=36.84$ MeV and dimensionless parameters $a_{0}=176.209, b_{0}=-250.992, c_{0}=100.253$.  For the upper limit and the lower limit~(the blue dashed lines in Fig.~\ref{eosfig}), $(a_1, b_1, c_1)$ = (1.55468, $-2.50096$, 1.44835) and (1.33832, $-2.0337$, 1.07001) respectively. 

For intermediate density, the stiff, intermediate and soft piecewise polytrope extension of the equation of states to higher densities from Ref.~\cite{Hebeler_2013} each with three exponents $\Gamma_1, \Gamma_2$ and $\Gamma_3$, can be written as follows,
\begin{eqnarray}
P(\rho) &=& \kappa_1 \rho ^{\Gamma_1}, \; \text{for} \; \rho_1 \leq \rho \leq \rho_{12},\notag \\
P(\rho) &=& \kappa_2 \rho ^{\Gamma_2}, \; \text{for} \; \rho_{12} \leq \rho \leq \rho_{23},\notag \\
P(\rho) &=& \kappa_3 \rho ^{\Gamma_3}, \; \text{for} \; \rho_{23}\leq \rho \leq \rho_{max}. \notag  \\  \label{eospp}
\end{eqnarray}
With mass density $\rho = m n$,
\begin{enumerate}
	\item the stiff EoS (red dashed line in Fig.~\ref{eosfig}) has the exponents $(\Gamma_1, \Gamma_2, \Gamma_3) = (4.5, 5.5, 3.0)$ where $(\rho_{12}, \rho_{23}, \rho_{max}) = (1.5\rho_s, 2.0\rho_s, 3.3\rho_s)$ and $(\kappa_1, \kappa_2, \kappa_3)$ = (11.6687, 51.7666, 2.56345).
	  
	\item the intermediate EoS has the exponents $(\Gamma_1, \Gamma_2, \Gamma_3) = (4.0, 3.0, 2.5)$ where $(\rho_{12}, \rho_{23}, \rho_{max}) = (3.0\rho_s, 4.5\rho_s, 5.4\rho_s)$ and $(\kappa_1, \kappa_2, \kappa_3)$ = (2.89711, 1.30607, 1.07402). 
	
	\item the soft EoS has the exponents $(\Gamma_1, \Gamma_2, \Gamma_3) = (1.5, 6.0, 3.0)$ where $(\rho_{12}, \rho_{23}, \rho_{max}) = (2.5\rho_s, 4.0\rho_s, 7.0\rho_s)$  and $(\kappa_1, \kappa_2, \kappa_3)$ = (0.0321845, 2.63607, 0.572502),
\end{enumerate} 
when the pressure and density are in the unit of $\text{GeV fm}^{-3}$ and $\text{MeV fm}^{-3}c^{-2}$ respectively. The density scale is $\rho_{s}c^{2}=150.273$~MeV$ $fm$^{-3}$.

These equations of states will be used to construct the $P-\mu$ diagram in Fig.~\ref{pmufig} for thermodynamic comparison with the multiquark phase.

\subsection{EoS of SS model for high density}

At high densities inside the NS core, baryons are tightly compressed, quarks and gluons would be so close together that individual quark and gluon are deconfined from a single baryon and yet the interaction could still be sufficiently strong. The gluons and quarks become deconfined but could still form bound states of multiquark. The multiquarks can possess colour charges in the deconfined phase while keeping the star colour singlet in totality, similar to ionized gas of positive and negative electric charges with total neutrality. In the multiquark model of Ref.~\cite{bch}, the colour charge is quantified by the number fractions of hanging radial strings $n_{s}$. For extreme density and low temperature~(less than a trillion Kelvin), the deconfined phase of quarks and gluons should be in the multiquark phase instead of the pure gas of weakly interacting quarks and gluons where perturbative QCD~(pQCD) is applicable. 

The multiquark EoS (\ref{eosmq1}),(\ref{eosmq2}) are expressed in dimensionless form. Apart from the colour charge parameter $n_{s}$, there is only one parameter we can choose to determine the entire behaviour of the EoS, the energy density scale $\epsilon_{s}$ which will give the physical density and pressure $\rho \epsilon_{s}, P\epsilon_{s}$. After choosing $\epsilon_{s}$, the corresponding distance scale of the SS model is fixed by $r_{0}=\left( G\epsilon_{s}/c^{4}\right)^{-1/2}$~\cite{bhp}.

The pQCD calculation, for the deconfined quarks and gluons of Ref.~\cite{Kurkela,Gorda} is also displayed for comparison in Fig.~\ref{eosfig}.

\subsection{Phase transition between confined baryonic matter and multiquark matter} \label{sectPT}

Under extreme pressure and density in low temperature environment~($T \lesssim 10^{12}$ K, the quark-gluon plasma phase transition temperature), baryons are compressed against one another so tightly that quarks inside begin to move freely among neighbouring baryons. Strong interaction is still strong and possibly nonperturbative. The baryonic matter should then undergo a deconfinement phase transition to the multiquark nuclear matter where the quarks form bound state via the remaining strong interaction in the deconfined vacuum~\cite{Bergman:2007wp,bch,bhp}. Following Ref.~\cite{Hoyos:2016zke}, we compare the free energy~(essentially negative pressure~\cite{Bergman:2007wp}) by {\it assuming} the {\it onset} value~($\mu_{0}=\mu(n=0)$) of chemical potential~(per quark) to be the same for baryonic matter and multiquark phase,
\be
\mu_{mq,0} = \frac{\mu_{b,0}}{3}N_{q}=\mu_{q,0}N_{q}, \label{mona}
\ee
where $N_{q}$ is the number of quarks in each multiquark and $\mu_{q,0}$ is the chemical potential per quark at the onset value. Using nuclear EoS, we set $\mu_{q,0}=308.55$ MeV~\cite{Hoyos:2016zke}. Since the SS model fixes $\mu_{mq,0}$ once the energy density scale $\epsilon_{s}$ is fixed, $N_{q}$ can be calculated subsequently. By this assumption, the phase transition is a first order since there will be a discontinuity in the density between nuclear matter and multiquark phase.

\begin{figure}
\centering
		\includegraphics[width=0.50\textwidth]{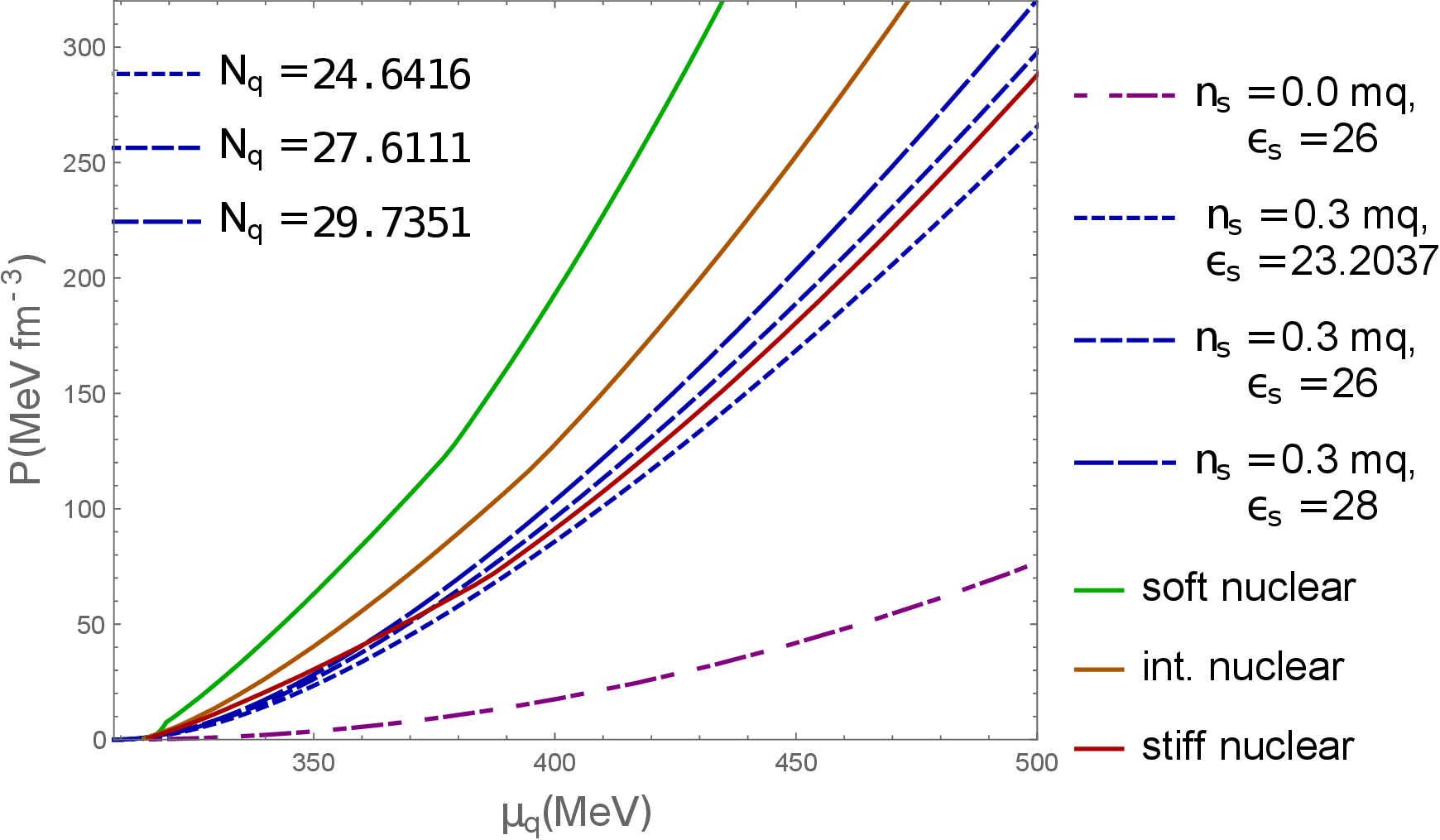}
		\caption{$P-\mu$ diagrams of multiquark comparing to stiff, intermediate and soft nuclear matter when the onset chemical potential value per quark is set to $308.55$ MeV. The energy density scale for the multiquark SS model is set to $\epsilon_{s}=23.2037, 26, 28$~GeV/fm$^{3}$~(three blue dashed lines) respectively.  }
		\label{pmufig}
\end{figure}

The transition point can be determined from the $P-\mu$ diagram as shown in Fig.~\ref{pmufig}. For chemical potential above the phase transition value, the pressure of the thermodynamically prefered phase will be larger. The multiquark EoS is presented with three choices of the energy density scale $\epsilon_{s}=23.2037, 26, 28$ GeV/fm$^{3}$ for $n_{s}=0, 0.3$. The colourless multiquark with $n_{s}=0$ is always less prefered thermodynamically than the nuclear EoS. The choices of $\epsilon_{s}$ are in the minimum range of values that would give NS masses within the constrained values from recent observations. As explained in subsequent section, the values of $\epsilon_{s}$ are also chosen so that the EoS interpolates reasonably well between the nuclear CET EoS at low densities and pQCD at high densities. 

Notably, the multiquark EoS with $\epsilon_{s}=23.2037, 26$ GeV/fm$^{3}$ are almost overlapping the stiff nuclear EoS of the CET. With $\epsilon_{s}=26$ GeV/fm$^{3}, n_{s}=0.3$, the multiquark phase is more prefered thermodynamically than the stiff nuclear EoS above the transition point at $\mu_{q}=374.0$ MeV.  For $\epsilon_{s}=26-28$ GeV/fm$^{3}$, the SS model predicts that in order to interpolate between the known stiff EoS at low density and the high density region consistent with conformal EoS, the core of massive NS should contain multiquark each composed of $N_{q}\simeq 25-30$ quarks corresponding to roughly $8-10$ baryons.

Fig.~\ref{pmufig} shows that the multiquark phase with $n_{s}=0$ is always less prefered thermodynamically than all of the nuclear equations of states and therefore we will not consider this possibility any further. On the other hand, the multiquark EoS for $n_{s}=0.3$ is notably almost identical to the stiff EoS of nuclear matter around the transition point demonstrating that it is a good extension of the stiff nuclear EoS to higher densities~(for $\epsilon_{s}\simeq 25$ GeV/fm$^{3}$, the multiquark and stiff EoS are overlapping almost completely). For intermediate and soft nuclear EoS, the multiquark phase is less prefered thermodynamically given the assumption (\ref{mona}). Here and henceforth we will consider only the possibility of the NS with multiquark core connecting to the {\it stiff} nuclear crust. 

For the aged neutron star where the temperature is low comparing to the chemical potential of the baryon and neutrino emission is suppressed, the charge neutrality and beta equilibrium are sustained. The direct Urca processes in the beta equilibrium are
\bea
n &\to & p + e +\bar{\nu}_{e}, \notag \\
p + e &\to & n + \nu_{e},
\ena
which implies
\be
\mu_{n}=\mu_{p}+\mu_{e}.
\ee
With strong gravitational pull, protons and electrons are confined within the star forming degenerate Fermi gases. As numbers of $p, e$ grow~(reduce), the Fermi energies increase~(decrease) and the number of $n, p, e$ saturate to constants at
\be
E_{F,n} = E_{F,p} + E_{F,e},   \label{fereqn}
\ee
where $E_{F,i}\simeq \mu_{i}$ is the Fermi energy of fermion $i=n, p, e$. Since $E_{F,i}\simeq n_{i}^{2/3}/m_{i}$ for low to moderate temperatures, Egn.~(\ref{fereqn}) dictates that the numbers of protons and electrons, $n_{p}=n_{e}$, are much less than the neutron. After the phase transition to the multiquark phase, the beta equilibrium is governed instead by
\be
E_{F,mq} = E_{F,mq'} + E_{F,e},   \label{fereqn1}
\ee
where $E_{F,mq}, E_{F,mq'}$ are the Fermi energy of the multiquark involved. As in (\ref{fereqn}), the charge neutrality demands $n_{mq'}=n_{e}$. Since the multiquarks are more massive than the baryon, roughly 8-10 times, while the number density drops by the same factor, i.e., $m_{mq}=f m_{b}, n_{mq}=n_{b}/f$, for $f\simeq 8-10$, the Fermi energy becomes $E_{F,mq}=f^{-5/3}E_{F,b}$, about 2-3\% of $E_{F,b}$. At beta equilibrium, the Fermi energy of electron has to reduce even further in order to satisfy (\ref{fereqn1}). Consequently, the number of electrons as well as the charged multiquarks is even more suppressed after the phase transition to the multiquark matter. Due to such electron suppression, the EoS of the multiquark is quite generic even with the consideration of the beta equilibrium.

\subsection{Matching of holographic multiquark EoS with low-density nuclear matter EoS}

\begin{figure}
		\centering
		{\includegraphics[width=0.49\textwidth]{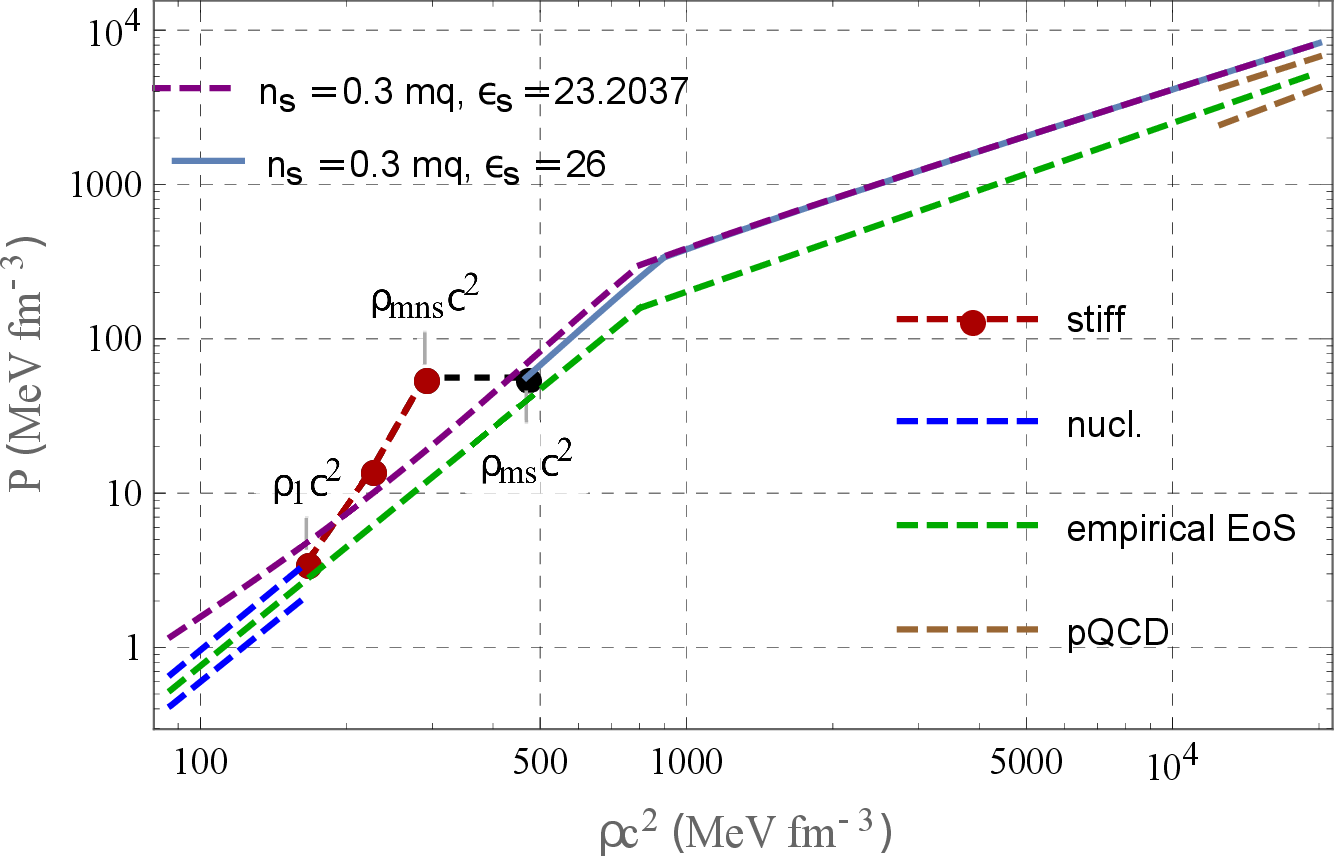} }
		\caption{EoS of multiquark interpolating between nuclear matter and extreme density region for $\epsilon_{s}=23.2037, 26$ GeV/fm$^{3}$, notice the density jump at the phase transition.}
		\label{eosfig}
	\end{figure}

The results, Fig. 1 and 2, of Ref.~\cite{Annala:2019puf} suggest a possibility that there is a double-power-law type EoS interpolating between the high and low density EoS given by pQCD and CET. One such candidate can be found in the early work of holographic SS model~\cite{bhp} where the multiquark phase is shown to dominate at large density and low temperature. By adjusting $\epsilon_{s}=23.2037$ GeV/fm$^{3}$ to give transition density $\rho_{c}c^{2}=0.8028$ GeV/fm$^{3}$ as suggested by the turning point of EoS in Fig. 1 of Ref.~\cite{Annala:2019puf}, a good interpolating equation of states of $n_{s}=0.3$ multiquark matter given by (\ref{eosmq1}), (\ref{eosmq2}) can be achieved as shown in Fig.~\ref{eosfig}. The green dashed line is the average empirical EoS connecting between pQCD and nuclear phases. As shown in Sect.~\ref{sectPT}, even though this EoS can interpolate well between low and high density, it is not thermodynamically prefered than the nuclear phases as suggested by CET. By increasing the energy density scale slightly to $\epsilon_{s}=26$ GeV/fm$^{3}$ to give transition density $\rho_{c}c^{2}=0.8996$ GeV/fm$^{3}$ as also depicted in Fig.~\ref{eosfig}, the multiquark phase becomes thermodynamically prefered than the stiff nuclear phase and still provide a good interpolation between low and high density. 

\section{MR diagram of NS with multiquark core} \label{sec-mr}

\begin{figure}[h]
		\centering
		\includegraphics[width=0.5\textwidth]{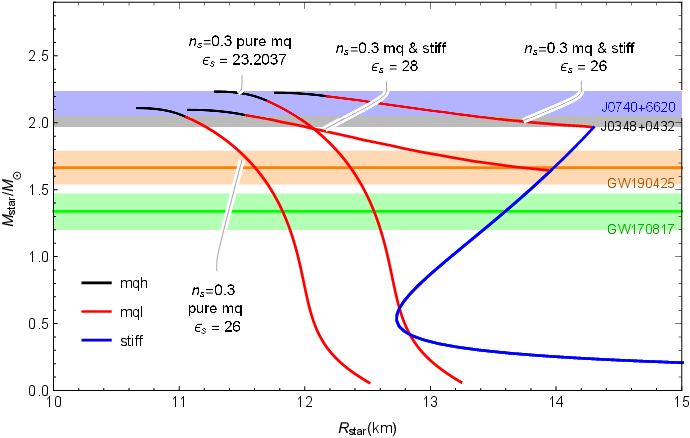}\\
		\includegraphics[width=0.5\textwidth]{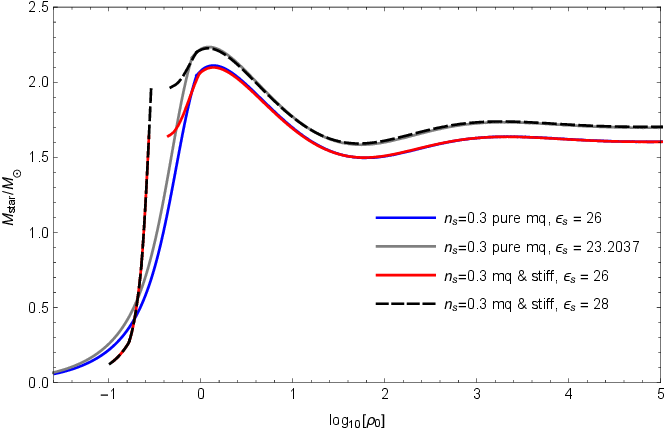}
		\caption{MR diagram and mass-central density of NS and quark star. The colour label represents the corresponding nuclear phase at the center of the star. Each point corresponds to a star with mass profile consisting of subsequent layers of nuclear phases in order of high to low density: multiquark, polytrope~(stiff), and CET. The pure hypothetical multiquark star has only multiquark layers. Observational NS masses are also presented for comparison.}		\label{mrfig}
	\end{figure}

The Tolman-Oppenheimer-Volkoff equation~\cite{tov1,tov2,bhp} is used in the consideration of mass profile of the NS,
\bea
\frac{dP}{dr}&&=-\frac{(\rho c^{2} +P)}{2}\frac{8\pi Pr^{3}+2M(r)}{r(r-2M(r))}, \notag \\
\frac{dM(r)}{dr}&&=4\pi \rho r^{2},
\ena
where $M(r)$ is the accumulated mass of the star up to radius $r$. In determination of the mass-radius diagram shown in Fig.~\ref{mrfig}, we use the multiquark EoS given in (\ref{eosmq1}),(\ref{eosmq2}) for high density region. As the density and pressure go down within the star and reach transition point with the stiff EoS, the new piecewise polytrope EoS (\ref{eospp}) is adopted until it reaches the low density region where the EoS given in (\ref{eq:P_Nucl}), (\ref{eq:E_Nucl}), and (\ref{EoS7}) will be used subsequently. From Fig.~\ref{eosfig}, we focus our consideration to 3 scenarios: \\

(i) $n_{s}=0.3,\epsilon_{s}=26~(28)$ GeV/fm$^{3}$ with transition to stiff at $\rho_{ms(2)}c^{2} =0.4678~(0.4389)$ GeV/fm$^3$ and $\rho_{mns(2)}c^{2} =0.2891~(0.2734)$ GeV/fm$^3$~(see Fig.~\ref{eosfig} where only $\epsilon_{s}=26$ GeV/fm$^{3}$ case is shown); \\

(ii) pure multiquark star with $n_{s}=0.3$ at $\epsilon_{s}=23.2037$ GeV/fm$^{3}$; \\

(iii) pure multiquark star with $n_{s}=0.3$ at $\epsilon_{s}=26$ GeV/fm$^{3}$. \\

The last two scenarios are the hypothetical multiquark star with no baryon crust. Scenario (ii) and (iii) are possible if the central temperature of the star is sufficiently high so that the surface temperature of the star is still higher than the nuclear-multiquark phase transition temperature.

From Fig.~\ref{mrfig} for NS containing multiquark core with $n_{s}=0.3$ continuing to stiff EoS, the maximum masses $\sim 2.2 M_{\odot}$ with radii around 11.8-11.1 km for $\epsilon_{s}=26-28$ GeV/fm$^{3}$, the larger energy density scale corresponds to smaller radius. For pure multiquark star with no baryon crust and $n_{s}=0.3$, the maximum mass for $\epsilon_{s}=23.2037~(26)$ GeV/fm$^3$ is $\sim 2.2~(2.1) M_{\odot}$ with radius around $11.3~(10.65)$ km respectively. An important prediction of the SS model is the existence of NS with multiquark core and stiff nuclear crust in the mass range $1.7-2.2 M_{\odot}$ and radii $14.5-11.1$ km  for $\epsilon_{s}=26-28$ GeV/fm$^{3}$ as depicted by the plateau-like black-red curve in the MR diagram of Fig.~\ref{mrfig}. For comparison, the NS masses with $1~\sigma$ uncertainties from observations~\cite{Antoniadis,Abbott:GW190425,Abbott:GW170817,Cromartie:2019kug} are also depicted in Fig.~\ref{mrfig}.

	\begin{figure}[h]
		\centering
		\includegraphics[width=0.5\textwidth]{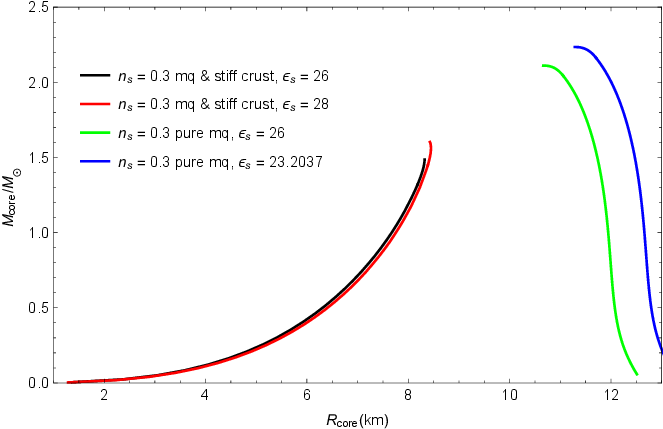}
		\caption{MR diagram of multiquark core for each curve in Figure~\ref{mrfig}. }		\label{mcrfig}
	\end{figure}

As shown in Fig.~\ref{mcrfig}, the multiquark {\it core} at the maximum mass for $n_{s}=0.3$ multiquark continuing to stiff EoS has mass and radius $1.49~(1.60) M_{\odot}$ and $8.3~(8.4)$ km for $\epsilon=26~(28)$ GeV/fm$^{3}$ respectively. For pure multiquark star, the mass and radius of the core are the same values as the entire star. Note that this multiquark {\it core} contains both high and low density layers governed by (\ref{eosmq2}) and (\ref{eosmq1}).

In order for the nuclear matter in the core of NS or the entire star to be in the multiquark phase, the central temperature and(or) density needs to be sufficiently large. The deconfinement phase transition temperature to quark-gluon plasma at low density around $n_{0}$ is estimated by the Heavy-Ion collisions at BNL RHIC~\cite{Arsene:2004fa} and CERN ALICE~\cite{ALICE:2017jyt} to be higher than $10^{12}$ K~(the Hagedorn temperature, $150$ MeV or $1.7 \times 10^{12}$ K). However at larger densities, theoretical models including the holographic SS model suggest the possibility that the QCD vacuum could become deconfined and quarks could form multiquark states while the chiral symmetry is still broken~\cite{Bergman:2007wp, bch} even at low temperature. Diquark in the colour superconductivity model~\cite{Alford:2007xm} and other multiquark~\cite{Jaffe:1976yi} could also form at high density and low temperature. Phase diagram in the SS model of Ref.~\cite{bch} shows the region of moderate temperature~($T<10^{12}$ K) and large $\mu$ where the multiquark in the deconfined vacuum is thermodynamically prefered down to the low temperature region. Moreover by the analysis in Section~\ref{sectPT}, the multiquark phase is more prefered thermodynamically than the confined stiff nuclear matter for sufficiently large $\mu_{q}$.  

 The temperature profile of the star can be calculated from the chemical potential profile within the star via the relation~(see e.g. Ref.~\cite{Burikham:2012kn}),
\be
\frac{T(r)}{T_{0}} = \frac{\mu(r)}{\mu_{0}}, \label{Tolr}
\ee
where $T_{0}, \mu_{0}=(\rho_{0}+P_{0})/n_{0}$ are the temperature and chemical potential at the reference point respectively. At the phase transition point, the deconfinement~(confinement) phase transition temperature between the multiquark and nuclear phase is determined by the chemical potential. For CET nuclear EoS, since the relevant Fermi energy is 36.84 MeV~\cite{Hebeler_2013}, the nuclear EoS is insensitive to temperature much below 1 MeV~(an estimation from Fig. 25 of Ref.~\cite{Holt:2014hma} gives approximately 5 \% increase in pressure for $T=1$ MeV from the zero temperature case). For $T<0.1$ MeV, the CET at zero temperature can thus be used in our determination of the transition chemical potential with an error less than 1 \%. By using (\ref{Tolr}), the transition values for the NS at $M_{max}$ with $n_{s}=0.3, \epsilon_{s}=26$ GeV/fm$^{3}$ multiquark core and stiff nuclear crust are $T_{\rm dec}=0.6741~T_{0}$ and $\mu_{\rm dec}=374.0$ MeV respectively. For the FYSS nuclear EoS~(see Section \ref{secTeff} for details), the transition occurs at $T_{\rm dec}\simeq 0.56~T_{0}$ and $\mu_{\rm dec}=341-342$ MeV as shown in Table~\ref{tab1}. For such hybrid star~(NS with multiquark core), the surface temperature is $T_{\rm surf}=0.5643~$(0.546-0.549)$~T_{0}$ for the CET~(FYSS) nuclear crust respectively. For a neutron star with surface temperature $10^{9}$ K, the core temperature would be around $1.77~(1.82-1.83)\times 10^{9}$ K for the CET~(FYSS) crust. 


Multiquark EoS contains two power-laws governing at high and low density, the corresponding multiquark matter is called the multiquark core and crust in Ref.~\cite{bhp}, but to avoid confusion we instead label them with ``mqh, mql''  respectively in this work. Each region gives different adiabatic indices $\gamma$ and sound speed $c_{s}$ as shown in Fig.~\ref{gamfig}. Interestingly, $\gamma \approx 1~(2.5)$ for high~(low) density multiquark respectively while $c_{s}^{2}>1/3$ violating the conformal bound for the high density region and most of the low density region. In the high density region $c_{s}^{2}\simeq 0.426$ for $n_{s}=0.3$, this is the value slightly above the conformal bound obeyed by the typical massless free quarks phase. The adiabatic index $\gamma$ of the high-density multiquark~(mqh) is very close to 1~(again the conformal limit of free quarks) while the low-density multiquark~(mql) has $\gamma \approx 2.5$, behaving more similar to the hadronic nuclear matter, but with colour charges and deconfined. On the other hand, $n_{s}=0$ colourless multiquark at high density has $\gamma \simeq 1.5, c_{s}^{2}\lesssim 0.55$. 

The maximum mass, corresponding radius, central density, and transition density for each variation of stars with multiquark cores are summarized in Table~\ref{tab1}.
\begin{table}
\footnotesize
\begin{tabular}{ |c|c|c|c|c|c| }
\hline
Matter content &$M_{max}$ & $R_{M_{max}}$& $\rho_{0}c^2$ & $\rho_{c}c^2$ & $\rho_{\text{mq\&b}}c^2$\\
inside the star &$(M_{\odot})$ & (km) & \scriptsize (GeV/fm$^{3}$) &\scriptsize (GeV/fm$^{3}$) &\scriptsize (GeV/fm$^{3}$) \\
\hline
&&&&& ($\rho_{\text{ms}}c^2$) \\
&&&&& 0.4678 \\
\scriptsize mq\&stiff &&&&& ($\rho_{\text{mns}}c^2$) \\
$ \epsilon_{s}$ = 26& 2.226  & 11.76 & 1.216 & 0.8996 & 0.2891\\\cline{6-6}
&&&&&$\mu_{\text{dec}}$(GeV)\\
&&&&& 0.3740\\
\hline
&&&&& ($\rho_{\text{ms2}}c^2$) \\
&&&&& 0.4389 \\
\scriptsize mq\&stiff &&&&& ($\rho_{\text{mns2}}c^2$) \\
$ \epsilon_{s}$ = 28 & 2.098  & 11.07 & 1.403 & 0.9688 & 0.2734 \\\cline{6-6}
&&&&&$\mu_{\text{dec}}$(GeV)\\
&&&&& 0.3605\\
\hline
&&&&& ($\rho_{\text{ms3}}c^2$) \\
\scriptsize mq\&FYSS, &&&&& 0.1246 \\
$i_Y = 1 \%$  &&&&& ($\rho_{\text{mns3}}c^2$) \\
$ \epsilon_{s}$ = 23.2037 & 2.235  & 11.31 & 1.246 & 0.8028 & 0.1111 \\\cline{6-6}
&&&&&$\mu_{\text{dec}}$(GeV)\\
&&&&& 0.3408\\
\hline
&&&&& ($\rho_{\text{ms4}}c^2$) \\
\scriptsize mq\&FYSS, &&&&& 0.1380 \\
$i_Y = 10 \%$  &&&&& ($\rho_{\text{mns4}}c^2$) \\
$ \epsilon_{s}$ = 23.2037 & 2.234  & 11.12 & 1.246 & 0.8028 & 0.1434 \\\cline{6-6}
&&&&&$\mu_{\text{dec}}$(GeV)\\
&&&&& 0.3424\\
\hline
\scriptsize pure mq &&&&&\\
$ \epsilon_{s}$ = 26& 2.111 & 10.66 & 1.396  & 0.8996 & - \\
&&&&&\\
\hline
\scriptsize pure mq&&&&&\\
$ \epsilon_{s}$ = 23.2067  & 2.235 & 11.29 & 1.246  & 0.8028 & - \\
&&&&&\\
\hline
\end{tabular}
\caption{Properties of massive neutron stars with multiquark cores ($n_s = 0.3$ and $\epsilon_{s}$ = 26, 28 GeV fm$^{-3}$) and stiff crust, massive neutron stars with multiquark cores ($n_s = 0.3$ and $\epsilon_{s}$ = 23.2037 GeV fm$^{-3}$) and nuclear matter crust with FYSS EoS in comparison with pure mutiquark stars ($n_s = 0.3$, $\epsilon_{s}$ = 26 GeV fm$^{-3}$ and $\epsilon_{s}$ = 23.2037 GeV fm$^{-3}$, respectively) at maximum masses.}
\label{tab1}
\end{table}

\subsection{Effects of finite temperature and the proton-baryon ratio on the MR diagram}  \label{secTeff}

\begin{figure}[h]
		\subfigure[$~n_{s}=0.3,\epsilon_{s}=23.2037$ GeV/fm$^{3}$ multiquark EoS and FYSS EoS. ]
		{\includegraphics[width=0.43\textwidth]{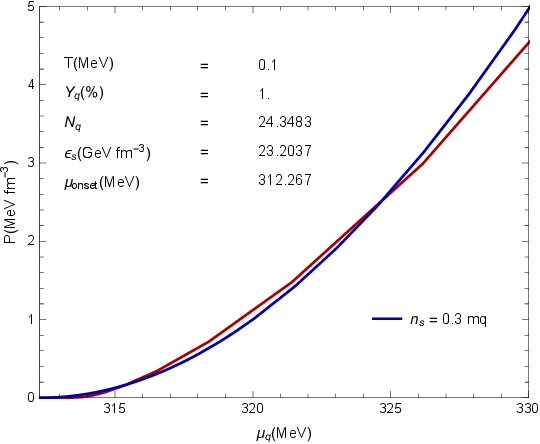} \label{pmut1}}
		\subfigure[$~n_{s}=0.3,\epsilon_{s}=23.2037$ GeV/fm$^{3}$ multiquark EoS and FYSS EoS. ]
		{\includegraphics[width=0.43\textwidth]{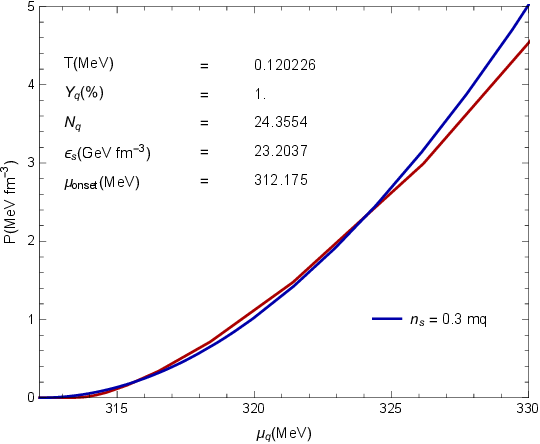}  \label{pmut2}}
		\subfigure[$~n_{s}=0.3,\epsilon_{s}=23.2037$ GeV/fm$^{3}$ multiquark EoS and FYSS EoS.]
		{\includegraphics[width=0.43\textwidth]{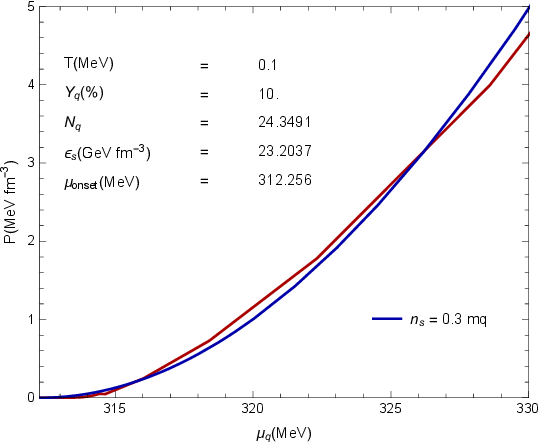}  \label{pmut3}}
		\caption{$P-\mu$ plots comparing the multiquark and baryonic nuclear phases at moderate temperatures and proton-baryon fractions.}
		\label{pmutfig}
\end{figure}

In this section, the effects of finite temperature and proton number density are explored. We will assume that the neutrino emissions have ceased and the entropy changes become negligible~(see e.g. Ref.~\cite{Prakash:1996xs,Pons:2001ar} for the generic analysis of protoneutron stars where the change of entropy is considered). For the baryonic nuclear matter crust, we adapt the FYSS EoS from Refs.~\cite{FYSS_2011,FYSS_2013,FYSS_2017}. The range of temperature for the FYSS EoS is $T=0.1-398$ MeV. For the multiquark the effect of temperature is negligible~(less than 0.4\%, see Ref.~\cite{bhp}). The differences between up and down quarks in a multiquark are assumed to be negligible and the multiquark EoS is independent of the proton-baryon ratio $Y_{q}\equiv n_{p}/n_{b}$. The $P-\mu$ analyses reveal that the multiquark phases with $\epsilon_{s}=26, 28$ are always thermodynamically prefered than the FYSS EoS. In Figure~\ref{pmutfig}, the $P-\mu$ plots for $\epsilon_{s}=23.2037$ GeVfm$^{-3}$ are presented for $T=0.10, 0.120226$ MeV and $Y_{q}=0.01, 0.10$.

\begin{figure}
		\centering
		{\includegraphics[width=0.49\textwidth]{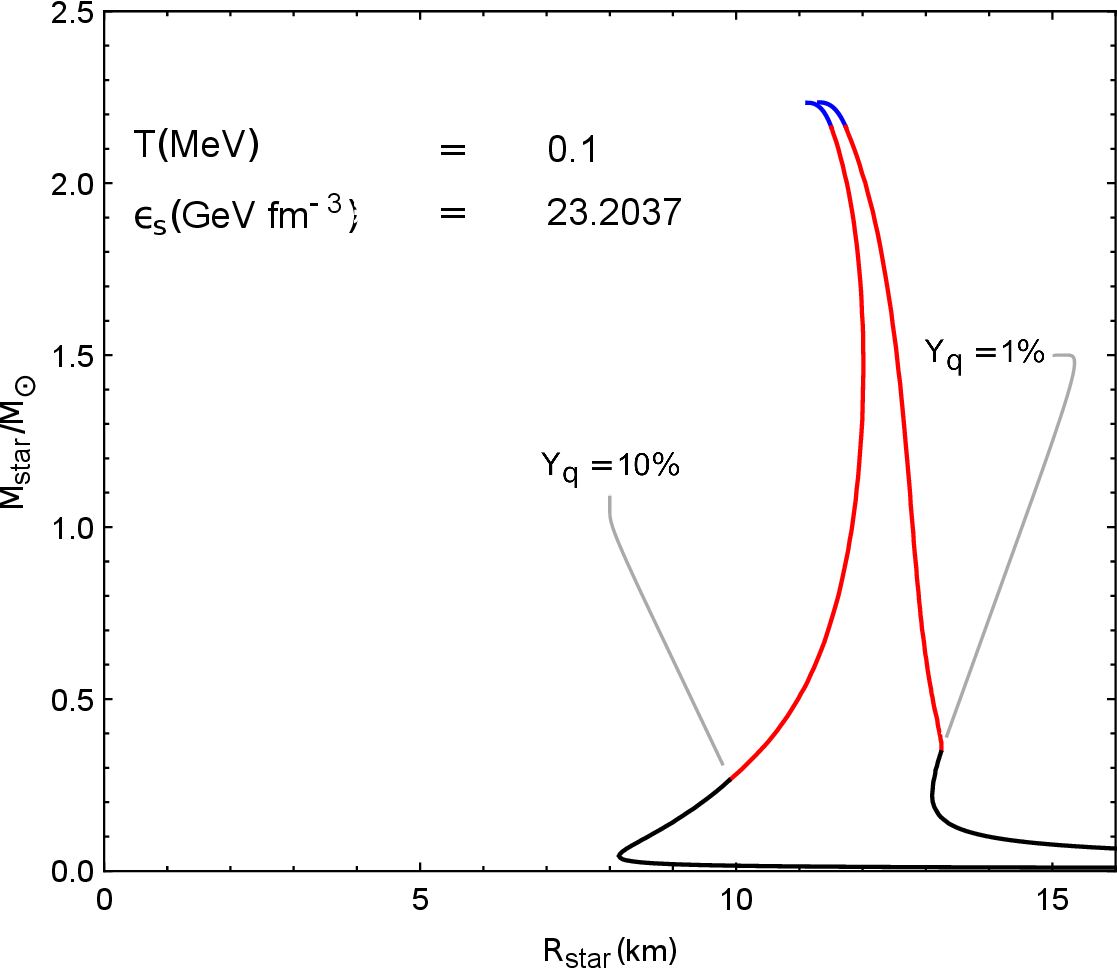} }
		\caption{MR diagram of NS with multiquark core and nuclear crust with FYSS EoS at $T=0.10$ MeV.}
		\label{MRTfig}
	\end{figure}

\begin{widetext}
	\begin{figure}[th]
		\centering
		\includegraphics[width=0.9\textwidth]{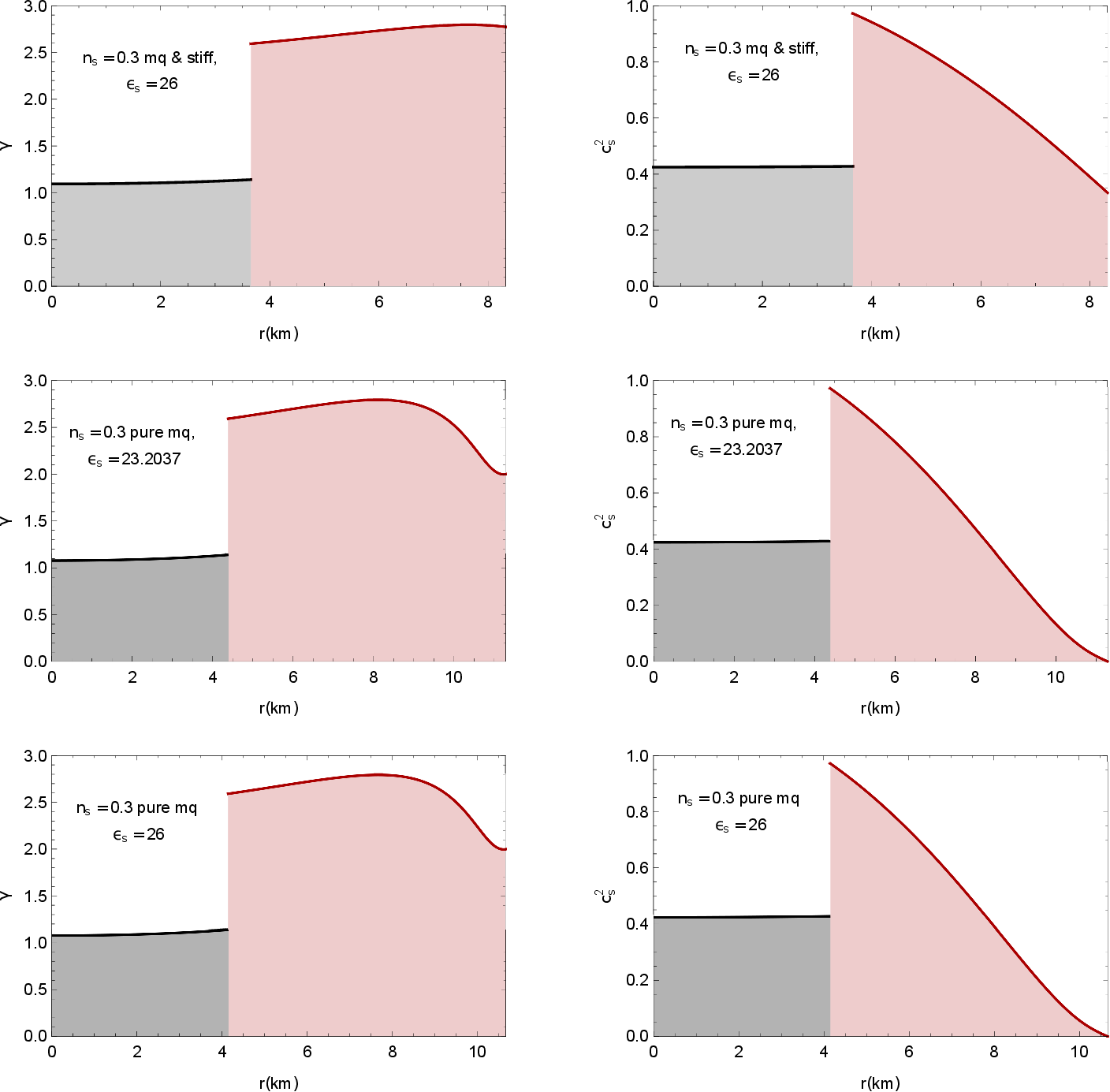}
		\caption{The adiabatic index $\gamma =\frac{d\ln P}{d\ln \rho}$ and $c_{s}^{2}$ of the multiquark {\it core} for each scenario, the multiquark core consists of two regions with high and low density given by (\ref{eosmq2}) and (\ref{eosmq1}). }		\label{gamfig}
	\end{figure}
\end{widetext}

The MR diagrams of NS with multiquark core for moderate temperatures and varying proton-baryon ratios are shown in Fig.~\ref{MRTfig}. Only $\epsilon_{s}=23.2037$ GeVfm$^{-3}$ are presented since for $\epsilon_{s}=26, 28$ GeVfm$^{-3}$, we would have only the {\it pure} multiquark star at such temperatures given the baryonic nuclear matter bahaves according to the FYSS EoS. The larger proton fraction shifts the radius of the NS with multiquark core to smaller values.

\section{Conclusions and Discussions}\label{sec-con}

The holographic SS model of multiquark nuclear matter has been applied to the inner core of NS at moderate to low temperature~(less than a trillion Kelvin). The EoS of the multiquark is interpolated between the high density and the low density where the CET is applicable. The transition density $\rho_{c}$ between the power laws in the empirical EoS is fixed once the energy density scale $\epsilon_{s}$ is chosen. The EoS of multiquark phase with colour-charge fraction $n_{s}=0.3$ can interpolate well between the high and low density regions when we set $\epsilon_s = 23.2037-28$ GeV/fm$^{3}$. This energy density scale corresponds to multiquark with the number of quarks $N_{q}\simeq 24-30$~(roughly $8-10$ baryons) per multiquark for $n_{s}=0.3$~(Fig.~\ref{pmufig}). Phase transitions from baryonic matter to deconfined multiquark have been studied and it is found that the multiquark phase at e.g. $n_{s}=0.3, \epsilon_s = 26, 28$ GeV/fm$^{3}$~(generically $>25$ GeV/fm$^{3}$) are more thermodynamically prefered than the stiff nuclear phase above the transition points.  

As shown in Fig.~\ref{eosfig}, the EoS for high-density multiquark has the same slope as the EoS of pQCD implying that its behavior could be more similar to that of free quark in spite of its bound state while the EoS for low-density multiquark passes through the region where the low-density nuclear EoS is used as good approximation. These nice behaviors imply that the existence of the multiquark phase~(which naturally contains high and low density profile as predicted by the SS model, see Ref.~\cite{bhp}) provides a missing link between the CET and the pQCD energy scales. MR diagram for various stiff-crust scenarios demonstrate that NS with multiquark core can have mass in the range $1.96-2.23~(1.64-2.10) M_{\odot}$ and radii $14.3-11.8~(14.0-11.1)$ km for $\epsilon_{s}=26~(28)$ GeV/fm$^{3}$ respectively. Note that the higher mass corresponds to smaller radius.  

At higher temperature in the order of few trillions Kelvin, the population of multiquarks should become less and the deconfined phase would consist mainly of weakly coupled quarks and gluons. Holographic models including the SS model predict the multiquark phase to be thermodynamically prefered than the QGP phase~\cite{bch} for moderate to low temperature at high densities. In newly formed NS or exotic quark star if the core temperature could reach over a few trillions Kelvin, it is possible to have this weakly-coupled quarks and gluons in the most inner core follow by multiquark layers resulting in even larger mass of the NS most likely larger than $2 M_{\odot}$. 

For aged NS with lower temperatures however, we expect only the multiquark phase to exist in the core. As density decreases with radial distance, the multiquark matter undergoes phase transition into confined baryonic matter or even coexist in mixed phase. For all scenarios that we consider, the NS with multiquark core could exist in a wide range of masses $M>2.0 M_{\odot}$ with radii around $11.1-14.3$ km for $n_{s}=0.3, \epsilon_{s}=26-28$ GeV/fm$^{3}$ and $11.1-12.1$ km for $n_{s}=0.3, \epsilon_{s}=23.2037$ GeV/fm$^{3}$ for the baryonic matter crust with FYSS EoS. The effects of moderate temperatures and proton-baryon fractions on the mass and radius of the NS are presented in Fig.~\ref{pmutfig}, \ref{MRTfig} where proton fractions in the baryonic crust lead to smaller NS at the same mass while the temperature effect is much less. There is a considerable number of observations of NS with masses above $2 M_{\odot}$, e.g. Ref.~\cite{Clark:2002db,Romani:2012rh,Romani:2012jf,vanKerkwijk:2010mt,Linares:2018ppq,Bhalerao:2012xe,Cromartie:2019kug,Nice:2005fi,Demorest:2010bx,Freire:2007sg,Quaintrell:2003pn}. It seems the massive NSs are abundant and our analyses suggest that they likely contain the multiquark cores. In Ref.~\cite{Abbott:2020lqk}, LIGO/Virgo set constraints on equatorial ellipticities of the millisecond pulsars to be less than $10^{-8}$. It would be interesting to explore the deformation of the NS containing multiquark core and check consistency with its EoS in the future work. However, the MR diagram of millisecond NS is minimally affected by the spin as demonstrated in Ref.~\cite{Miller:2019cac}.

\begin{acknowledgments}
P.B. is supported in part by the Thailand Research Fund (TRF),
Office of Higher Education Commission (OHEC) and Chulalongkorn University under grant RSA6180002. S.P. is supported in part by the Second Century Fund: C2F PhD Scholarship, Chulalongkorn University.

\end{acknowledgments}

\end{document}